%Paper: hep-th/9405075
%From: sobczyk@evalvx.ific.uv.es
%Date: 11 May 94 19:29:00 WET DST
%Date (revised): 7 Jun 94 09:37:00 WET DST

\input harvmac
\rightline{FTUV/94-25}
\rightline{hep-th/9405075}
\rightline{revised}
\vskip 2truecm
\centerline{\bf CLASSICAL $r$-MATRICES}
\vskip .5truecm
\centerline{\bf AND CONSTRUCTION OF QUANTUM GROUPS}
\vskip 2truecm
\centerline{J.Sobczyk\foot{permanent address (after June 15th): Institute
of Theoretical Physics, Wroc\l aw University, Pl. Maxa Borna 9, 50-205,
Wroc\l aw, Poland; e-mail: jsobczyk@plwruw11.bitnet}
}
\vskip .5truecm
\centerline{Departamento de Fisica Teorica, Facultad de Fisica}
\centerline{Universidad de Valencia, E-46100 Burjassot, Valencia, Spain}
\vfill
\noindent
\centerline{ABSTRACT}
\medskip
A problem of constructing quantum groups from classical $r$-matrices is
discussed.
\vfill
\vskip 2truecm\noindent May 1994.

\eject
1. Classical $r$-matrices
\ref\dra{V.G.Drinfeld, Soviet. Math. Dokl. {\bf 27} (1983) 68}
\ref\oth{J.-H.Lu and A.Weinstein, J. Diff. Geometry {\bf 31} (1990) 501;
M.A.Semenov-Tian-Shansky, Publ. RIMS, Kyoto University, {\bf 21} (1985) 1237}
play a very important role in a theory of quantum
groups
\ref\rtf{N.Yu.Reshetikhin, L.A.Takhtadzyan and L.D.Faddeev, Leningrad Math. J.,
{\bf 1} (1990) 193},\
\ref\drb{V.G.Drinfeld, {\sl Quantum groups}, Proc. ICM, Berkeley, 1986,
vol.1, 789}.
They are closely related to a structure of Poisson-Lie group which
appears as a classical limit of a quantum group. Poisson-Lie group is a Lie
group $G$ together with a Poisson bracket defined on it in such a way that
the group action becomes a Hamiltonian map. It is convenient to define the
Poisson bracket by means of a bivector field $\Lambda$
\eqn\bra{
\{ \phi , \psi\} = \Lambda (d\phi , d\psi )
}
or rather to consider instead of $\Lambda$ an object $\eta :G \rightarrow
{\cal G}\otimes {\cal G}$ (${\cal G}$ is a Lie algebra of $G$)
\eqn\el{
\eta (x) = (T_{\rho^{-1}_x}\otimes T_{\rho^{-1}_x})_x \Lambda_x
}
where $\rho$ denotes a right group action. The fact that the action
of the group is a Hamiltonian map can be expressed as a cocycle condition on
$\eta$
\eqn\coc{
\eta (g_1g_2) = \eta (g_1) + {\rm Ad}^{\otimes} (g_1) \eta (g_2)
}
Possible solutions of this equation are in the form of coboundary
\eqn\cob{
\eta (g) = {\rm Ad}^{\otimes} r - r
}
where $r\in {\cal G}\otimes {\cal G}$ is called a classical $r$-matrix.
A requirement that the Poisson bracket defined by $\Lambda$ via $r$ satisfies
the Jacobi identities imposes some conditions on $r$. Namely $r$
has to satisfy so called modified classical Yang-Baxter equation (MCYBE)
which means that the following object constructed out of $r$ (Schouten
bracket)
\eqn\sch{
<r, r>\equiv [r_{12}, r_{13}] + [r_{12}, r_{23}] + [r_{13}, r_{23}]
}
is an invariant tensor (under the adjoint action) in ${\cal G}\otimes
{\cal G}\otimes {\cal G}$.

As it is relatively easy to find solutions of \sch\ one can construct a lot of
examples of Poisson-Lie groups. Moreover for many interesting
groups every $\eta (g)$
is a coboundary. This is e.g. the situation of complex simple Lie groups \dra ,
Lorentz group
\ref\zaa{S.Zakrzewski, {\sl Poisson structures on the Lorentz group},
to appear in Lett. Math. Phys.}
and the group of symmetries of a
space-time of arbitrary metric signature in dimensions $\geq 3$
\ref\zab{S.Zakrzewski, {\sl Talk at the Karpacz Winter School of Theoretical
Physics} 1994, unpublished}. Two
dimensions are exceptional in this sense as for the euclidean group in $D=2$
it is
possible to construct $\eta (g)$ which is not a coboundary. One should have
in mind that also
other examples are known in which the structure of Poisson-Lie group
is not given by a classical $r$ matrix \ref\szz{
I.Szymczak and S.Zakrzewski, J.Geom.Phys., {\bf 7} (1990) 553}.
\bigskip\bigskip

2. It is
clear that by studying classical $r$-matrices one gets some insight into a
possible structure of quantum groups. For example
it is known that classifications
of $r$-matrices for D=4 Lorentz and Poincar\' e groups gives similar patterns
to that of the analogous classification of quantum deformations of those
groups \zaa , \zab ,
\ref\zaw{S.L.Woronowicz and S.Zakrzewski, Compositio Mathematica {\bf 90}
(1994)
211},
\ref\pod{P.Podle\' s, {\sl Talk at the Karpacz Winter School of Theoretical
Physics} 1994, unpublished}.

In this paper the starting point is a $r$-matrix satisfying
MCYBE. We shall try to investigate what can be said about (possible)
existing quantum groups having a Poisson-Lie group defined by this $r$ as
its classical limit. In other terms we shall try to construct a quantum
group out of $r$ only. This problem has been discussed by several authors
\ref\gur{
see e.g. D.Gurewicz, V.Rubtsov and N.Zobin, J.Geom.Phys. {\bf 9} (1992) 25,
S.Zakrzewski, {\sl Geometric quantization of Poisson groups -- diagonal and
soft deformations}, Contemp. Math. 1993
}.
Our discussion will be on much more elementary level.

One possible way of reasoning is the following. With the help of $r$ one can
construct Poisson bracket in the space of functions on $G$ (we shall always
have in mind elements of some suitable chosen matrix representation of $G$).
Then one can try to replace Poisson brackets by commutators. However
naive it looks like this is a method to construct some new examples of
quantum groups: $\kappa$-Poincar\'e
\ref\zac{S.Zakrzewski, {\sl Quantum Poincar\' e group related to $\kappa$-
Poincar\' e algebra}, J. Phys. {\bf A27} (1994) 2075},
ISL(2,{\bf C}) group \ref\mas{P.Ma\'slanka, {\sl The quantum ISL(2,{\bf C})
group}, \L\'od\'z\ University preprint, IMUL 9/93}
and supergroups: $\kappa$-superPoincar\'e
\ref\klms{P.Kosi\' nski, J.Lukierski, P.Ma\' slanka and J.Sobczyk,
{\sl Quantum deformation of the Poincar\` e supergroup and $\kappa$-deformed
superspace}, Wroc\l aw University preprint IFTWr 868/94}.
In general it is clear that the described method is ambiguous due to
appearance of operator ordering problems. Only in some particular cases these
ambiguties are not present or can be relatively easy solved. In the
case of $sl(2)$ ($[H, X]=2X,\ \ [H, Y]=-2Y,\ \  [X, Y]=H$, group
element is $g=\pmatrix{a&b\cr c&d}$)
with $r$-matrix $r=X\wedge Y$ we obtain a typical Poisson
bracket $\{a , b\} = ab$ and it is a priori difficult to guess that an
ordering leading to a "good"
quantum group is $[a, b] = \lambda ab + (1-\lambda )
ba$ with $q= {2-\lambda \over 1-\lambda}$ with similar or even more complicated
tricks for other commutators. Very similar is the situation with Poisson
brackets coming from $r$ matrix $r=X\wedge H$.

This is a proper place to comment on the notion of a "good"
quantum group . At the beginning there is some matrix group.
Quantum group is a Hopf algebra with elements
polynomials constructed out of noncommutative
matrix elements modulo some relations among them.
Usually these relations are introduced by $R$ matrix (the notation is
explained in \rtf )
\eqn\rel{
RT_1T_2=T_2T_1R
}
but in general it need not to be a case. The comultiplication defined
on the algebra generators is the same way as in the
classical (undeformed) case (an advantage of defining quantum relations by
means of \rel\ is that comultiplication becomes then the Hopf algebra
homomorphism).
Important is that both deformed and undeformed (i.e. commutative)
algebras of polynomials should be of the same size i.e. that
the dimensionality of the space of "deformed" polynomials in matrix elements
should be the same as in the undeformed case. Moreover there should exist
a deformation
parameter such that in the limit when it goes to zero both algebras become
identical. Usually it is taken as granted
that the statement about the dimensionalities
is satisfied if relations are introduced by a $R$ matrix satisfying the quantum
Yang Baxter equation (QYBE):
\eqn\ybe{
R_{12} R_{13} R_{23} = R_{23} R_{13} R_{12}
}
But it turns out to be nontrivial to prove it even in simple cases
\ref\wor{S.Woronowicz, Rep. Math. Phys. {\bf 30} (1991) 259}.
Similarly one can demand also that
\eqn\nowe{
\tilde R R=\lambda {\bf 1}\otimes
{\bf 1}
}
(here $\tilde R=PRP$ where $P(x\otimes y)=y\otimes x$).
We should however remember that the famous $R$ for $sl(n)$ discussed in
\rtf\ satisfies \ybe\ but not \nowe .
We shall come back to this point later.

Let us also notice that when we discuss
different choices of normal ordering in the
case of $sl(2)$ what is wrong about almost all orderings is that they
give rise to too restrictive quadratic relations.
\bigskip\bigskip

3. Going beyond the naive quantization scheme it is known that in some cases
one can construct out of $r$ an element $R={\rm e}^{hr}$ satisfying QYBE.
As $R={\bf 1}\otimes {\bf 1}+hr+...$ it is clear that
$r$ has to satisfy CYBE, and not only the MCYBE (it means that the expression
given in \sch\    equals zero). Among
examples that can be treated in this way let us mention $sl(2)$ with
$r$-matrix $r=X\wedge H$ in the matrix representation
$X=\pmatrix{0&1\cr 0&0},\ \ Y=\pmatrix{0&0\cr 1&0},\ \ H=\pmatrix{1&0\cr
0&-1}$
\ref\zad{S.Zakrzewski, Lett. Math. Phys., {\bf 22}, (1991) 287}.
In this way the nonstandard deformation of $sl(2)$ has been obtained.
Another example is provided by a Poincar\' e group. Here group elements
are described by 5-dimensional matrices $g=\pmatrix{\Lambda &x\cr 0&1}$.
Poincar\'e algebra elements are represented
by matrices $P_{\mu}=e_{\mu}^4,\ \
M_j=\epsilon_{jkl} e_l^k,\ \ L_j = e_0^j + e_j^0$ and r matrix is
$r=M_3\wedge L_3\ +\ \alpha P_1\wedge P_2
\ +\ \beta P_0 \wedge P_3$.
This $r$-matrix is taken from the list given in \zab . It satisfies CYBE
but that does not imply that $R=e^{hr}$ has to
satisfy QYBE. It can however be calculated that
$$
R_{12}R_{13}R_{23}=R_{23}R_{13}R_{12}={\bf 1}\otimes {\bf 1}\otimes {\bf 1} +$$
$$+\alpha (P_1\otimes P_2\otimes 1- P_2\otimes P_1\otimes 1 +
P_1\otimes 1\otimes P_2 - P_2\otimes 1\otimes P_1 +
1\otimes P_1\otimes P_2 - 1\otimes P_2\otimes P_1) + $$
$$+\beta (P_0\otimes P_3\otimes 1- P_3\otimes P_0\otimes 1 +
P_0\otimes 1\otimes P_3 - P_3\otimes 1\otimes P_0 +
1\otimes P_0\otimes P_3 - 1\otimes P_3\otimes P_0)$$
$$+sin(h)(M_3\otimes L_3\otimes  1 -L_3\otimes M_3\otimes 1 +
M_3\otimes 1\otimes L_3 - L_3\otimes 1\otimes M_3 +$$
$$+1\otimes M_3\otimes L_3 - 1\otimes L_3\otimes M_3) +
(1-cos (h))(M_3^2\otimes L_3^2\otimes {\bf 1} +
L_3^2\otimes M_3^2\otimes {\bf 1}  +$$
$$+M_3^2\otimes 1\otimes  L_3^2 +   L_3^2\otimes 1\otimes M_3^2 +
1\otimes M_3^2\otimes L_3^2 + 1\otimes L_3^2\otimes M_3^2) +$$
$$+\alpha sin(h) (P_2\otimes L_3\otimes P_2 + P_1\otimes L_3\otimes P_1) +$$
$$+\beta sin (h) (P_0\otimes M_3\otimes P_0 - P_3\otimes M_3\otimes P_3)+$$
$$+\alpha (1 - cos (h))(P_2\otimes L_3^2\otimes P_1 -
P_1\otimes L_3^2\otimes P_2)+$$
$$+\beta (1 - cos (h))(P_0\otimes M_3^2\otimes P_3 -
P_3\otimes M_3^2\otimes P_0)+$$
$$+sin^2(h)(M_3^2\otimes L_3\otimes L_3 - M_3\otimes L_3^2\otimes M_3 +
L_3^2\otimes M_3\otimes M_3+$$
$$- L_3\otimes M_3^2\otimes L_3 +
M_3\otimes M_3\otimes L_3^2 + L_3\otimes L_3\otimes M_3^2)+$$
$$+(1-cos(h))^2(-M_3^2\otimes L_3^2\otimes 1_4 - L_3^2\otimes 1_4\otimes
M_3^2 - 1_4\otimes M_3^2\otimes L_3^2)+$$
$$+sin(h)(1-cos(h))(-M_3\otimes L_3\otimes 1_4 + L_3\otimes M_3\otimes 1_4$$
\eqn\long{
-M_3\otimes 1_4\otimes L_3 + 1_4\otimes L_3\otimes M_3 +
L_3\otimes 1_4\otimes M_3 - 1_4\otimes M_3\otimes L_3)
}
In the above formula $1_4$ denotes a diagonal matrix with 1's in the first
four rows.
\bigskip\bigskip

4. Interesting enough there are some examples in which $R=e^{hr}$
does not satisfy
QYBE however when put into \rel\
gives rise to a good (established by means of other methods)
quantum group $G$. This is the situation of the
mentioned above $\kappa$-Poincare group with $r=\sum_j L_j\wedge P_j$ \zac .
Similar is the situation of $D=2$ euclidean group
in three-dimensional representation ($X=\pmatrix{0&0&1\cr 0&0&0\cr 0&0&0}$
$J=\pmatrix{0&1&0\cr -1&0&0\cr 0&0&0}$)
with the $r$-matrix $r=J\wedge X$
\ref\wla{A.Ballesteros, E.Celeghini, R.Giachetti, E.Sorace and M.Tarlini,
{\sl An R-matrix approach to the quantization of the euclidean group E(2)},
to be published in Journal of Physics A; P.Ma\' slanka, J. Math. Phys.,
{\bf 35} (1994) 1976}.
Still another example is provided by the Heisenberg group with
its Lie algebra generators represented as $A=\pmatrix{0&1&0\cr 0&0&0\cr
0&0&0}$,
\ $A^+=\pmatrix{0&0&0\cr 0&0&1\cr 0&0&0}$,\ \ $H=\pmatrix{0&0&1\cr 0&0&0\cr
0&0&0}$ and $r=A\wedge A^+$
\ref\wlb{E.Celeghini, R.Giachetti, E.Sorace and M.Tarlini, J. Math. Phys.,
{\bf 32} (1991) 1155}.

As this point can give rise to some confusion let us comment on it
shortly. The fact that $R$ satisfies or not QYBE has nothing to do with
whether the resulting Hopf algebra is associative or not. It is always assumed
to be associative. QYBE is just a constraint on $R$ that is supposed
to ensure that
the space of cubic (and then higher orders) polynomials in noncommuting
matrix elements variables can be large enough
to be of the same dimensionality as in the undeformed case
(it eliminates a possibility of extra unwanted cubic relations).
Of course QYBE is not a necessary
condition to be so. Suppose for a moment that $R$ does not satisfy the
QYBE. Then it is possible to define an invertible element $B$ by means of
\eqn\byb{
R_{12}R_{13}R_{23} = B R_{23}R_{13}R_{12}.
}
$B$ has necessarly the properties (they follow from the fact
that the algebra is associative)
\eqn\bin{
t_{ij}t_{kl}t_{mn}B_{ns\ lr\ jp} = B_{mn\ kl\ ij} t_{jp}t_{lr}t_{ns}
}
As we work in the particular representation in which $T$ is given
$B$ is a 3-fold tensor product of numerical matrices.
$t_{ij}$ are matrix elements which commutation relations are given
in \rel . $B$ can be viewed as a generalization of the element $<r,r>$ in
\sch\ to the case of noncommuting $t_{ij}$. If we analyze its structure
as a power series in the deformation parameter $h$ we get
\eqn\be{
B={\bf 1}\otimes{\bf 1}\otimes{\bf 1} + h^2<r,r> + ...
}

In a very similar way if the relations are introduced by a $R$ matrix which
does not satisfy the condition \nowe\
we automatically obtain an
invertible element $C=\tilde R R$ with the properties
\eqn\ce{
C_{ij\ kl} t_{jm} t_{ln}=t_{ij}t_{kl}C_{jm\ ln}
}

Both conditions: \nowe\ and \ybe\ are desirable for construction of good
quantum groups but certainly neither is a necessary one
(strictly speaking it is even
not obvious if they are sufficient; one should have in
mind that for general matrix groups the resulting relations are not purely
quadratic -- e.g. for the Poincar\' e group, another complication comes from
extra relations given e.g. by the determinant).
It would be interesting to
find some weaker conditions on $R$. Equations \bin\ and \ce\ are
however not suitable for that purpose as they are identities in the algebra
defined by \rel\ and so carry no information about the space of polynomials
the size of which we would like to be able to estimate.
\bigskip\bigskip

5. It turns out to be interesting to analyze in more detail the case of
$R=e^{hr}$ where $h$ is a deformation parameter
and $r$ is a classical antisymmetric $r$ matrix. Such $R$ have two interesting
general properties. First of all they  satisfy \nowe .
Moreover they give rise to Poisson-Lie group structure on $G$ in
the limit as $h\rightarrow 0$. What is missing is an estimate of the
dimensionality of the space of elements of the
resulting Hopf algebra. We do not know if the
requirement of the proper dimensionality
can be expressed as some condition on $R$. We have checked only
that in some simple examples one quite
surprisingly obtains correct results.

The first such example is given by $sl(2)$ with $r$-matrix
$r=X\wedge Y$. In the given above two-dimensional representation
for $X$ and $Y$ one calculates that $R={\rm e}^{hr}$ equals
\eqn\rsl{
R={\bf 1}\otimes {\bf 1} + {\rm sin}(h)(X\otimes Y - Y\otimes X)
+ ({\rm cos}(h) -1)(U\otimes D + D\otimes U)
}
where $U=\pmatrix{1&0\cr 0&0}$ and $D=\pmatrix{0&0\cr 0&1}$. This $R$ does
not satisfy QYBE. When however put into \rel\ it gives rise to the standard
deformation of $sl(2)$ with $q={{\rm cos}(h)\over 1+{\rm sin}(h)}$.

As we explained above as a byproduct of this construction
of a standard deformation of $sl(2)$
we are able to
calculate an element $B$ with the interesting property
\bin . In the explicite way it is
given as
$$B={\bf 1}\otimes {\bf 1}\otimes {\bf 1} +
(-sin^4(h)+sin^5(h))U\otimes U\otimes D
+(-sin^4(h)-sin^6(h))U\otimes D\otimes U$$
$$+(-sin^4(h)+sin^5(h))U\otimes D\otimes D + (-sin^4(h)-sin^5(h))
D\otimes U\otimes U$$
$$+(-sin^4(h)-sin^6(h))D\otimes U\otimes D + (-sin^4(h) - sin^5(h))
D\otimes D\otimes U$$
$$+cos(h)(sin^2(h)+sin^3(h)+sin^4(h))X\otimes Y\otimes U$$
$$+cos(h)(-sin^2(h)+sin^3(h)+sin^5(h)) X\otimes Y\otimes D$$
$$+cos(h)(-sin^2(h)-sin^3(h) -sin^5(h)) Y\otimes X\otimes U$$
$$+cos(h)(sin^2(h)-sin^3(h)+sin^4(h)) Y\otimes X\otimes D$$
$$+(-sin^2(h)+sin^4(h))X\otimes U\otimes Y + (sin^2(h)-sin^6(h))
X\otimes D\otimes Y$$
$$+(sin^2(h)-sin^6(h))Y\otimes U\otimes X + (-sin^2(h) +sin^4(h))
Y\otimes D\otimes X$$
$$+cos(h)(sin^2(h)-sin^3(h)+sin^4(h))U\otimes X\otimes Y$$
$$+cos(h)(-sin^2(h)-sin^3(h)-sin^5(h))D\otimes X\otimes Y$$
$$+cos(h)(-sin^2(h)+sin^3(h)+sin^5(h))U\otimes Y\otimes X$$
\eqn\bsl{
+cos(h)(sin^2(h)+sin^3(h)+sin^4(h))D\otimes Y\otimes X
}

Also the two-parameter deformation of the $gl(2)$ can be obtained in
this way. Here the starting point is $r=\alpha X\wedge Y
+ \beta U\wedge D$ (of course $U={H+1 \over 2}$ and $D={1-H \over 2}$). It
is possible to calculate that with $s=h\sqrt{ \beta^2-\alpha^2}$
\eqn\sch{
R={\bf 1}\otimes {\bf 1}+ (ch(s)-1)(U\otimes D + D\otimes U)+
}
$$+{sh(s) \over \sqrt{ \beta^2-\alpha^2}} (\alpha X\wedge Y + \beta U\wedge
D)$$
Using this R (which does not satisfy QYBE) one obtains the
standard relations: $ab=qba,\ ac=pca,\ cd=qdc,\ bd=pdb,\ qbc=pcb,\
[a,d]=(p-{1\over q})cb=(q-{1\over p})bc$
with
\eqn\prr{
q={
\sqrt{\beta^2 - \alpha^2} ch(s)+\beta sh(s) \over
\sqrt{\beta^2 - \alpha^2}
+\alpha sh(s)
}, \ \ p={
\sqrt{\beta^2 - \alpha^2} - \alpha sh(s) \over
\sqrt{\beta^2 - \alpha^2} ch(s) + \beta sh(s)}
}
In the limit $\beta \rightarrow 0$ one gets $p\rightarrow q$.

It would be in our opinion interesting to continue the study of
properties of $R$ matrices of the form $R=e^{hr}$. That includes both a
discussion of new examples and investigation a general
question what are the sufficient conditions on $r$
to give rise to a good quantum group.

\vskip 2truecm
\noindent
ACKNOWLEDGMENTS.

I wish to thank dr. P.Ma\'slanka for interesting  discussions on
sl(2). I thank dr. S.Zakrzewski for pointing out errors in the first
version of this paper and for many stimulating comments about the notion
of a good (being of correct size) quantum group. Finally I would like to
thank prof. J. de Azc\' arraga for a warm
hospitality at Valencia where this work was done. The author was supported
by EEC grant number ERBCIPACT920488.
\listrefs
\end